\documentclass[conference]{IEEEtran}
\IEEEoverridecommandlockouts

\usepackage{cite}
\usepackage{amsmath,amssymb,amsfonts}
\usepackage{algorithmic}
\usepackage{graphicx}
\usepackage{textcomp}
\usepackage{xcolor}
\usepackage{booktabs}
\usepackage{multirow}
\usepackage{url}

\def\BibTeX{{\rm B\kern-.05em{\sc i\kern-.025em b}\kern-.08em
    T\kern-.1667em\lower.7ex\hbox{E}\kern-.125emX}}

\begin{document}

\title{Format-Controlled Multi-Scale JPEG Compression\\Response Analysis for Image-Level Compression-History Forgery Screening}

\author{
  \IEEEauthorblockN{Sujith K Mandala}
    \IEEEauthorblockA{sujithkmandala15@gmail.com}
}

\maketitle

\begin{abstract}
Image-level screening for JPEG compression-history mismatch is a practical subtask in digital image forensics, especially when large image collections must be triaged before heavier pixel-level localization or human forensic review.
We propose a lightweight, interpretable feature engineering pipeline for image-level compression-history forgery screening using only CPU computation and gradient boosted trees.
Our method introduces \emph{multi-scale Error Level Analysis} (ELA) computed at seven JPEG quality levels, combined with novel \emph{cross-quality ELA ratio} features that capture double-compression artifacts characteristic of spliced regions, augmented by spatial entropy, FFT energy bands, edge density, SRM residuals, and DCT blockiness, yielding a 405-dimensional feature vector.
CASIA v2.0 contains a format confound (60\% of tampered images are TIFF while authentic images are JPEG/BMP and contain no TIFF samples), enabling a trivial \texttt{is\_tiff} classifier to reach 0.80 AUC.
We address this through rigorous format-controlled evaluation: on the JPEG-only subset (9,501 images, eliminating the TIFF/JPEG container confound), our method achieves AUC~=~0.990 [95\% CI: 0.988--0.991] and F1~=~0.905 using 5-fold stratified cross-validation.
Under a conservative source-aware group split that prevents images sharing parsed source identifiers from crossing folds, AUC remains 0.976.
An ablation study reveals that multi-scale ELA provides the dominant gain (+0.180 AUC over single-quality on the format-controlled subset), while cross-quality ratios provide complementary double-compression detection.
These results support that the method detects compression-history inconsistencies rather than file-format shortcuts---while offering feature-level interpretability, CPU-only deployment, and sub-second inference.
External transfer tests on Columbia splicing blocks (AUC = 0.708) and CoMoFoD copy-move as a negative control (AUC = 0.499, chance level) delimit the method's operating scope to compression-history mismatch detection rather than universal image manipulation detection.
The proposed method should therefore be interpreted as a JPEG compression-response screening tool, not as a universal detector for all image manipulation mechanisms.
\end{abstract}


\section{Introduction}
\label{sec:intro}

The proliferation of sophisticated image editing tools has made digital image forgery increasingly accessible, posing serious threats to journalism, legal proceedings, identity verification systems, and social media integrity~\cite{farid2009image}.
Common manipulation operations---including region splicing, copy-move duplication, and content-aware removal---leave subtle but detectable traces in the statistical properties of an image's pixel values and compression artifacts.

Deep learning approaches such as ManTra-Net~\cite{mantranet2019}, MVSS-Net~\cite{mvss2022}, and RGB-N~\cite{rgbn2018} have achieved strong detection and localization performance.
However, these methods are typically GPU-accelerated, contain millions of trainable parameters, and often lack interpretability---characteristics that limit their practical deployment in forensic workflows where auditability and computational efficiency are paramount.

Error Level Analysis (ELA)~\cite{krawetz2007a} is a classical technique that exploits JPEG compression artifacts: when an image is re-saved at a given quality factor $q$, authentic regions converge to a uniform error level while forged regions (which underwent a different compression history) exhibit distinctive residual patterns.
Despite its intuitive appeal, single-quality ELA suffers from sensitivity to the unknown original compression quality and produces unreliable results across diverse image sources.

In this work, we address these fundamental limitations through three key innovations:
\begin{enumerate}
    \item \textbf{Multi-scale ELA:} Computing ELA at seven quality levels ($q \in \{30, 50, 60, 75, 80, 90, 95\}$) captures compression artifacts across the full quality spectrum, making the features robust regardless of the original image's compression history.
    \item \textbf{Cross-quality ELA ratios:} By computing pairwise ratios between ELA maps at different quality levels, we derive features that are invariant to absolute residual magnitude and directly capture the double-JPEG compression signatures characteristic of forgery.
    \item \textbf{Complementary feature families:} We augment ELA features with spatial entropy analysis, frequency-domain features (FFT), edge density statistics, SRM steganalysis residuals~\cite{fridrich2012rich}, and DCT blockiness measures, creating a comprehensive 405-dimensional feature vector.
\end{enumerate}

These features are classified using a HistGradientBoosting classifier~\cite{sklearn}, which runs without GPU acceleration and processes each image in under 0.5 seconds.

A critical methodological concern with CASIA v2.0, potentially under-controlled in prior CASIA-based evaluations, is that 60\% of tampered images are stored as TIFF while all authentic images are JPEG/BMP, with no TIFF samples among authentics.
A trivial \texttt{is\_tiff} classifier achieves 0.80 AUC, and file metadata alone reaches 0.92 AUC.
We conduct rigorous format-controlled experiments: evaluating on only the JPEG subset (7,437 authentic + 2,064 tampered), our method achieves AUC = 0.990 [95\% CI: 0.988--0.991], supporting that it detects compression-history inconsistencies rather than file-format shortcuts.
Under a conservative source-aware group split based on recoverable CASIA filename identifiers, AUC remains 0.976.

\textbf{Operating scope.}
The proposed method is designed for image-level screening of JPEG compression-history inconsistencies.
It is not intended to replace pixel-level localization methods or copy-move-specific correspondence detectors.
Its strongest operating regime is the detection of manipulations that introduce heterogeneous compression or recompression traces, such as cross-source splicing or heterogeneously processed composites.

The remainder of this paper is organized as follows: Section~\ref{sec:related} reviews related work in image forensics. Section~\ref{sec:method} details our feature extraction pipeline and classification approach. Section~\ref{sec:experiments} presents experimental results including ablation studies and computational benchmarks. Section~\ref{sec:discussion} discusses findings and limitations, and Section~\ref{sec:conclusion} concludes with future directions.

\section{Related Work}
\label{sec:related}

\subsection{Compression-Based Forensic Features}

JPEG compression introduces characteristic blocking artifacts at $8\times8$ pixel boundaries and quantization effects in the DCT domain.
Ye et al.~\cite{ye2007detecting} exploited DCT coefficient discontinuities to detect double JPEG compression.
Lin et al.~\cite{lin2009fast} proposed a fast algorithm for detecting JPEG artifacts using the blocking artifact matrix.
These methods demonstrate that compression artifacts carry rich forensic information, but typically focus on a single compression quality factor.

Krawetz~\cite{krawetz2007a} popularized Error Level Analysis as a visual forensic tool, computing the pixel-wise difference between an image and its re-compressed version at a fixed quality.
While intuitive, single-quality ELA is unreliable when the original quality factor is unknown---a limitation we explicitly address through multi-scale analysis.

\subsection{Statistical and Filter-Based Methods}

Fridrich and Kodovsk\'{y}~\cite{fridrich2012rich} introduced rich models for steganalysis using co-occurrence statistics of high-pass filter residuals (SRM).
These noise-domain features were later adapted for splicing and copy-move detection~\cite{cozzolino2015splicebuster}.
Pan et al.~\cite{pan2012exposing} used local noise variance inconsistency to expose splicing boundaries.
Popescu and Farid~\cite{popescu2004exposing} exploited periodic artifacts introduced by geometric resampling.

Our work incorporates SRM residuals as one of seven feature families, complementing them with compression-specific features that target different manipulation traces.

\subsection{Deep Learning Approaches}

Bayar and Stamm~\cite{bayar2016deep} proposed constrained convolutional layers that learn manipulation-detecting filters.
ManTra-Net~\cite{mantranet2019} introduced an anomaly detection framework for manipulation tracing without requiring pixel-level labels during training.
MVSS-Net~\cite{mvss2022} achieved strong localization performance using multi-view multi-scale supervision.
RGB-N~\cite{rgbn2018} combined RGB stream features with noise-sensitive features for joint detection and localization.

While these methods achieve impressive performance, they are typically GPU-accelerated in practical deployments, contain millions of trainable parameters, and provide limited interpretability regarding which manipulation traces trigger detection.
Our approach targets deployment scenarios where these constraints are prohibitive, and we additionally contribute a format-controlled evaluation methodology that is not always explicit in CASIA v2.0-based evaluations.

\section{Methodology}
\label{sec:method}

\begin{figure}[t]
    \centering
    \includegraphics[width=\columnwidth]{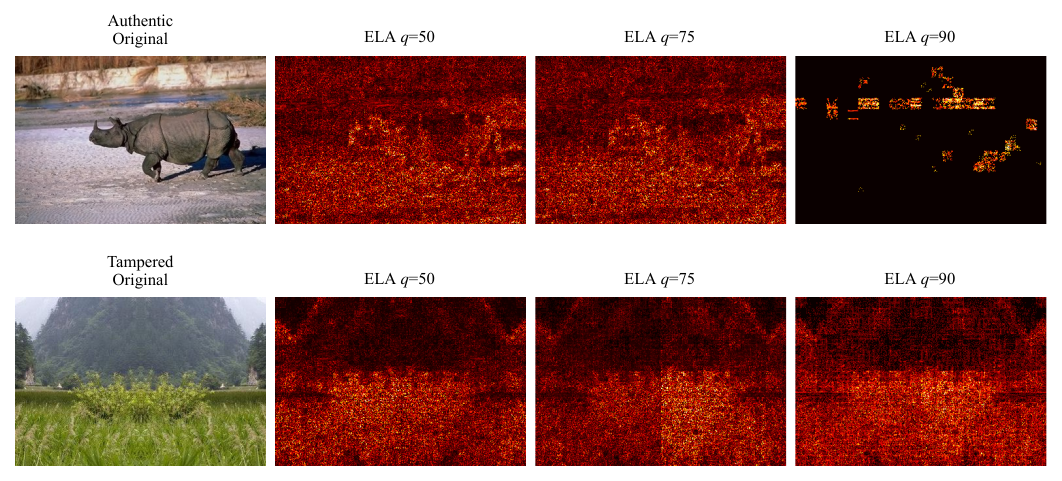}
    \caption{ELA visualization for an authentic image (top) and tampered image (bottom) at quality levels 50, 75, and 90. The tampered image shows inconsistent residual patterns in the manipulated region across quality levels.}
    \label{fig:ela_vis}
\end{figure}

Our pipeline consists of three stages: (1) multi-family feature extraction from raw images, (2) feature combination and normalization, and (3) gradient-boosted classification.
Fig.~\ref{fig:ela_vis} illustrates the ELA visualization at multiple quality levels for authentic and tampered examples.

\subsection{Preprocessing}

Given an input image $I$, we convert to RGB color space and resize such that the maximum dimension does not exceed 1024 pixels (preserving aspect ratio via Lanczos interpolation).
This ensures consistent feature extraction across diverse image resolutions while preserving sufficient detail for artifact analysis.

\subsection{Feature Family 1: Multi-Scale ELA (266 features)}

For each of seven JPEG quality levels $q \in Q = \{30, 50, 60, 75, 80, 90, 95\}$, we compute the ELA residual map:
\begin{equation}
    E_q(x,y,c) = \left| I(x,y,c) - \hat{I}_q(x,y,c) \right|
    \label{eq:ela}
\end{equation}
where $\hat{I}_q$ denotes image $I$ re-saved at JPEG quality $q$ and reloaded, and $c \in \{R,G,B\}$ denotes the color channel.

From each ELA map $E_q$, we extract:
\begin{itemize}
    \item \textbf{Global statistics} (6 features per quality): mean, standard deviation, 75th percentile, 95th percentile, 99th percentile, and the fraction of pixels exceeding $\mu + 2\sigma$ (high-residual fraction).
    \item \textbf{Spatial grid statistics} (32 features per quality): The image is divided into a $4\times4$ grid, and for each cell we compute the mean and standard deviation of residuals, capturing spatial heterogeneity.
\end{itemize}

This yields $7 \times (6 + 32) = 266$ ELA features.
The rationale for multiple quality levels is that different forgery types leave artifacts at different compression scales: high-quality splicing is most visible at low $q$ (where re-quantization amplifies subtle differences), while low-quality paste operations are detectable at high $q$ (where authentic regions have already converged).

\subsection{Feature Family 2: Cross-Quality ELA Ratios (16 features)}

The key insight motivating this feature family is that double-JPEG compression produces characteristic ratio signatures across quality levels.
When a region has been previously saved at quality $q_1$ and the composite image is later re-saved at $q_2 \neq q_1$, the ratio of ELA residuals at two different analysis qualities differs systematically from single-compression regions.

We compute channel-averaged ELA maps $\bar{E}_q(x,y) = \frac{1}{3}\sum_c E_q(x,y,c)$ and derive ratio maps:
\begin{equation}
    R_{q_a,q_b}(x,y) = \frac{\bar{E}_{q_a}(x,y)}{\bar{E}_{q_b}(x,y) + \epsilon}
    \label{eq:ratio}
\end{equation}
where $\epsilon = 10^{-6}$ prevents division by zero.

For four quality pairs $(q_a, q_b) \in \{(30,90), (50,90), (60,80), (30,60)\}$, we extract the mean, standard deviation, 5th percentile, and 95th percentile of each ratio map, yielding 16 features.
These pairs span both wide-gap ratios (30/90) that amplify double-compression effects and narrow-gap ratios (60/80) that capture subtle quality-factor mismatches.

\subsection{Feature Family 3: ELA Entropy (18 features)}

Spatial entropy of the ELA map captures the information-theoretic heterogeneity introduced by splicing.
In authentic images, ELA residuals have relatively uniform entropy across spatial regions; forged images exhibit localized entropy anomalies at manipulation boundaries.

Using the $q=50$ ELA map (normalized to $[0, 255]$ and quantized), we divide the image into a $4\times4$ grid and compute Shannon entropy per cell:
\begin{equation}
    H_k = -\sum_{b=1}^{32} p_{k,b} \log_2(p_{k,b} + \epsilon)
    \label{eq:entropy}
\end{equation}
where $p_{k,b}$ is the normalized 32-bin histogram of cell $k$.
We extract all 16 cell entropies plus their standard deviation and range, yielding 18 features.

\subsection{Feature Family 4: FFT Radial Energy Bands (6 features)}

Manipulation operations often alter the frequency distribution in ways not detected by spatial analysis.
We compute the 2D DFT of the grayscale image (resized to $256\times256$) and partition the centered magnitude spectrum into five concentric radial bands: $[0,16)$, $[16,32)$, $[32,64)$, $[64,96)$, $[96,128)$ pixels from center.
For each band, we compute the fractional energy.
Additionally, we compute the high-to-low frequency energy ratio ($E_{r\geq64} / E_{r<32}$).
Splicing often introduces high-frequency boundary artifacts visible in these band features.

\subsection{Feature Family 5: Edge Density (20 features)}

Forgery boundaries introduce spatial discontinuities detectable via edge analysis.
We compute the Sobel gradient magnitude on the grayscale image (normalized to $[0,1]$):
\begin{equation}
    S(x,y) = \sqrt{S_x(x,y)^2 + S_y(x,y)^2}
\end{equation}
We extract: global mean, standard deviation, and 95th percentile (3); a $4\times4$ grid of mean edge density (16); and cross-cell standard deviation (1).

\subsection{Feature Family 6: SRM Steganalysis Residuals (27 features)}

Following Fridrich and Kodovsk\'{y}~\cite{fridrich2012rich}, we apply three $5\times5$ high-pass SRM kernels to each RGB channel of the image (resized to $224\times224$).
For each of the 9 filter-channel combinations, we extract the mean, standard deviation, and 95th percentile of the absolute response.

\subsection{Feature Family 7: DCT Blockiness, Noise, and Color (52 features)}

This family combines:
\textbf{DCT blockiness} (8): boundary-to-interior intensity ratios and their spatial variance.
\textbf{Laplacian noise} (32): variance and mean absolute Laplacian in a $4\times4$ grid.
\textbf{Color statistics} (12): inter-channel correlations and per-channel standard deviation, skewness, and kurtosis.

\subsection{Classification}

The complete 405-dimensional feature vector is classified using HistGradientBoostingClassifier~\cite{sklearn} with: 1000 maximum iterations with early stopping (patience 50), learning rate 0.03, maximum depth 8, minimum 20 samples per leaf, L2 regularization 1.0, and 255 histogram bins.
Class imbalance is addressed via inverse-frequency sample weighting.

\subsection{Computational Complexity}

The dominant cost is seven JPEG re-compression operations, each $O(n)$ where $n$ is pixel count.
Total feature extraction is $O(n)$ per image.
Classification is $O(d \cdot T \cdot \log B)$ where $d=405$, $T$ is boosting iterations, and $B=255$ bins---negligible compared to extraction.

\section{Experiments}
\label{sec:experiments}

\subsection{Dataset}

We evaluate on CASIA v2.0~\cite{dong2013casia}, containing:
\begin{itemize}
    \item \textbf{7,491 authentic images} (7,437 JPEG + 54 BMP) across 8 scene categories.
    \item \textbf{5,123 tampered images} (2,064 JPEG + 3,059 TIFF) with splicing (\texttt{Tp\_S\_}) and copy-move (\texttt{Tp\_D\_}) manipulations.
\end{itemize}
Total: 12,614 images, tampered rate 40.6\%.

\textbf{Format confound.}
A significant bias exists: 60\% of tampered images are TIFF while all authentic images are JPEG/BMP.
This means non-forensic metadata features (file extension, file size, resolution) can achieve up to 0.92 AUC via format shortcuts (Table~\ref{tab:shortcut}).
Prior results on mixed-format CASIA v2.0 may be susceptible to format-related confounding unless such factors are explicitly controlled.
We address this through multiple controlled evaluation protocols described below.

\subsection{Evaluation Metrics}

We report AUC (threshold-independent), F1 at validation-selected threshold (median across folds), and classification accuracy.
All results use out-of-fold predictions aggregated over five stratified folds with fixed random seed 42.
F1 thresholds are selected \emph{per-fold on validation data} and the median threshold is applied globally, preventing threshold overfitting.

\subsection{Format Leakage Assessment}

Table~\ref{tab:shortcut} quantifies the format confound using non-forensic metadata features only.

\begin{table}[t]
\centering
\caption{Shortcut baseline AUC using only file metadata (no image content). These quantify dataset bias considered in the controlled protocols below.}
\label{tab:shortcut}
\begin{tabular}{@{}lcc@{}}
\toprule
Metadata Feature & AUC & F1 \\
\midrule
\texttt{is\_tiff} flag only & 0.798 & 0.748 \\
Extension (one-hot) & 0.800 & 0.748 \\
Width + height + aspect ratio + file size & 0.919 & 0.840 \\
All metadata combined & 0.924 & 0.848 \\
\bottomrule
\end{tabular}
\end{table}

These results demonstrate that any method evaluated on the full mixed-format CASIA v2.0 dataset without controlling for format may be susceptible to format-related confounding.
Our pipeline explicitly does \emph{not} use filename, extension, EXIF metadata, or file size as input features---but the TIFF images differ from JPEG in their compression response to ELA, creating an indirect confound that we must evaluate against.

\subsection{Main Results: Format-Controlled Evaluation}

Our primary result uses the \textbf{JPEG-only subset} (7,437 authentic JPEG + 2,064 tampered JPEG = 9,501 images), which eliminates the TIFF/JPEG container confound because all retained images share the same file container.

\begin{table}[t]
\centering
\caption{Format-controlled evaluation on CASIA v2.0. JPEG-only is the primary result; other protocols provide additional validation.}
\label{tab:main}
\begin{tabular}{@{}lccc@{}}
\toprule
Evaluation Protocol & AUC & F1 & Acc. \\
\midrule
JPEG-only subset (9,501 imgs) & \textbf{0.990} & \textbf{0.905} & \textbf{0.958} \\
\quad 95\% CI & [0.988--0.991] & --- & --- \\
Grouped JPEG-only (source-aware) & 0.976 & 0.840 & --- \\
All$\rightarrow$JPEG q95 normalization & 0.952 & 0.869 & 0.884 \\
All$\rightarrow$JPEG q85 normalization & 0.835 & 0.747 & 0.760 \\
All$\rightarrow$JPEG q75 normalization & 0.842 & 0.753 & 0.768 \\
Mixed format (all 12,614 imgs) & 0.971 & 0.897 & 0.909 \\
\bottomrule
\end{tabular}
\end{table}

Key observations:
\begin{itemize}
    \item The JPEG-only AUC (0.990) is \emph{higher} than the mixed-format result (0.971), indicating that TIFF images---which lack JPEG compression history---were actually \emph{harder} for our ELA-based method (reducing performance when included).
    \item Under a conservative source-aware group split that prevents images sharing parsed source identifiers from crossing folds, AUC remains 0.976.
    \item Converting all images to JPEG q95 before extraction (removing file-container and decoder-level differences, though not fully equalizing prior compression history) yields 0.952 AUC on the full dataset.
    \item The q85/q75 normalization (AUC 0.835/0.842) shows expected degradation: aggressive recompression erases the compression-history traces the method relies upon. This is consistent with a method that detects compression-response inconsistencies. The small difference between q85 and q75 is not interpreted as meaningful; both settings show substantial degradation under aggressive recompression.
\end{itemize}

Per-fold AUC for the JPEG-only evaluation: 0.991, 0.991, 0.990, 0.990, 0.989 ($\sigma = 0.001$), demonstrating exceptional stability.

\textbf{Source-aware grouping protocol.}
The grouped split uses source identities recoverable from CASIA v2.0 filenames.
Authentic images are assigned a source identity from the category-index pattern, e.g.,
\texttt{Au\_ani\_00018} $\rightarrow$ \texttt{ani00018}.
For tampered images, we parse all recoverable CASIA-style source identifiers embedded in the filename.
When multiple identifiers are present, we construct a source graph in which identifiers appearing in the same tampered filename are connected.
Group identifiers are then defined by connected components of this graph, so that images sharing any parsed donor or target identity are assigned to the same GroupKFold partition when those identities are recoverable from filenames.
For samples where additional donor or target identities are not explicitly recoverable, residual source-overlap risk may remain.
This yields 897 unique source groups assigned into 5 deterministic GroupKFold partitions.
We release the parsed source-group file, fold assignments, and out-of-fold prediction files to support independent verification of the grouping protocol.

\subsection{Comparison with Prior Work}

Table~\ref{tab:comparison} contextualizes our results.
We separate same-protocol baselines (our implementation, identical data/splits) from literature-reported numbers that use different metrics (pixel-level localization AUC), different CASIA subsets, and different supervision settings.

\begin{table}[t]
\centering
\caption{Comparison with same-protocol baselines and literature context on CASIA-related benchmarks. $^\dagger$GPU-accelerated. $^\ddagger$Published localization/detection AUC from prior papers; not directly comparable to our image-level JPEG-only protocol.}
\label{tab:comparison}
\begin{tabular}{@{}lccc@{}}
\toprule
Method & AUC & F1 & Hardware \\
\midrule
\multicolumn{4}{@{}l}{\textit{Same-protocol baselines (our implementation, 5-fold CV, JPEG-only):}} \\
\quad ELA (single $q$=75) & 0.806 & 0.595 & CPU \\
\quad Cross-quality ratios only & 0.985 & 0.882 & CPU \\
\quad Multi-scale ELA only & 0.986 & 0.892 & CPU \\
\midrule
\multicolumn{4}{@{}l}{\textit{Literature context$^\ddagger$ (published AUC, mixed protocols and varying splits):}} \\
\quad CFA1~\cite{ferrara2012image} & 0.522 & --- & CPU \\
\quad RGB-N$^\dagger$~\cite{rgbn2018} & 0.795 & --- & GPU \\
\quad ManTra-Net$^\dagger$~\cite{mantranet2019} & 0.817 & --- & GPU \\
\midrule
\textbf{Ours (JPEG-only, image-level)} & \textbf{0.990} & \textbf{0.905} & \textbf{CPU} \\
\bottomrule
\end{tabular}
\end{table}

\noindent\textbf{Note:} Published AUC values are copied from the benchmark tables in~\cite{rgbn2018} and~\cite{mantranet2019}; method citations indicate original descriptions.
These values use different CASIA subsets, train/test splits, supervision settings, and pixel-level localization metrics, and are therefore included only as literature context.
Our primary contribution is the format-controlled methodology; the same-protocol baselines above provide the meaningful ablative comparison.
Additional lightweight same-protocol feature-family baselines are reported in Table~\ref{tab:families}, including DCT/noise/color (0.831 AUC), SRM residuals (0.712 AUC), edge density (0.736 AUC), ELA entropy (0.741 AUC), and FFT energy bands (0.615 AUC), all under the same JPEG-only protocol.

\subsection{Ablation Study}

Table~\ref{tab:ablation} shows the incremental contribution of each feature family.

\begin{table}[t]
\centering
\caption{Ablation: incremental feature addition (JPEG-only subset).}
\label{tab:ablation}
\begin{tabular}{@{}lccr@{}}
\toprule
Feature Configuration & AUC & F1 & \#Feat \\
\midrule
Base ELA ($q=75$ only) & 0.806 & 0.595 & 38 \\
Multi-scale ELA (7 levels) & 0.986 & 0.892 & 266 \\
\quad + Cross-quality ratios & 0.989 & 0.902 & 282 \\
\quad + ELA entropy & 0.989 & 0.905 & 300 \\
\quad + FFT energy bands & 0.990 & 0.905 & 306 \\
\quad + Edge density & 0.990 & \textbf{0.906} & 326 \\
\quad + SRM + DCT + noise + color & \textbf{0.990} & 0.905 & 405 \\
\bottomrule
\end{tabular}
\end{table}

Key findings:
\begin{itemize}
    \item \textbf{Multi-scale ELA is dominant:} +0.180 AUC (0.806 $\rightarrow$ 0.986) from extending to seven quality levels. This is the single most important design decision.
    \item \textbf{Cross-quality ratios add discrimination:} +0.003 AUC from only 16 features that encode double-compression signatures.
    \item \textbf{Complementary families provide marginal, metric-dependent changes:} Remaining families contribute +0.001 AUC collectively (0.989 $\rightarrow$ 0.990), suggesting redundancy but possible robustness benefits.
\end{itemize}

The ablation suggests that the method's power derives primarily from multi-scale compression response analysis, rather than from any single auxiliary feature family.

\subsection{Individual Feature Family Performance}

Table~\ref{tab:families} shows each family used in isolation.

\begin{figure}[t]
    \centering
    \includegraphics[width=\columnwidth]{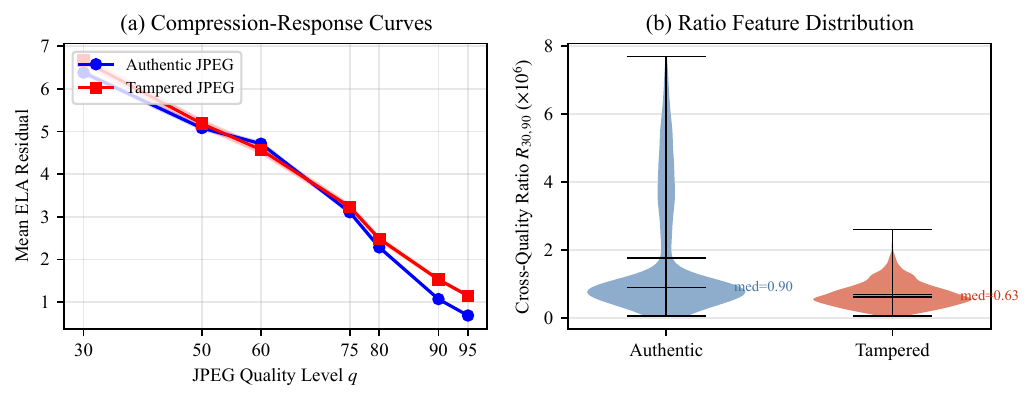}
    \caption{(a) Mean ELA residual across JPEG quality levels for authentic vs.\ tampered images (JPEG-only subset). Tampered and authentic images exhibit different compression-response profiles, especially at high quality factors. (b) Violin plot of the cross-quality ratio $R_{30,90}$ (mean, $\times 10^6$) showing class-dependent distributional shift from only 16 ratio features (standalone AUC = 0.985).}
    \label{fig:response}
\end{figure}

\begin{table}[t]
\centering
\caption{Individual feature family performance (JPEG-only subset).}
\label{tab:families}
\begin{tabular}{@{}lcr@{}}
\toprule
Feature Family & AUC & \#Features \\
\midrule
Multi-scale ELA & \textbf{0.986} & 266 \\
Cross-quality ratios & 0.985 & 16 \\
DCT + noise + color & 0.831 & 52 \\
ELA entropy & 0.741 & 18 \\
Edge density & 0.736 & 20 \\
SRM residuals & 0.712 & 27 \\
FFT energy bands & 0.615 & 6 \\
\bottomrule
\end{tabular}
\end{table}

Multi-scale ELA alone achieves 0.986 AUC on the JPEG-only subset, indicating that compression artifacts carry the dominant forensic signal.
Remarkably, the 16 cross-quality ratio features alone reach 0.985 AUC, demonstrating that relative compression behavior across quality levels is nearly as discriminative as full ELA maps.
DCT/noise/color features (0.831) provide moderate independent performance, while entropy, edges, SRM, and FFT contribute weaker but complementary signals.

\textbf{Feature-level analysis.}
Univariate screening on the JPEG-only subset reveals that the top-10 most discriminative individual features are all ELA spatial variability measures at $q=95$ (best: \texttt{ela\_q95\_std}, AUC = 0.855).
No single feature exceeds 0.90 AUC individually, yet the 16 cross-quality ratio features---which individually peak at only 0.72 AUC---collectively achieve 0.985 AUC when combined, demonstrating strong synergistic interaction.

\subsection{Computational Efficiency}

\begin{table}[t]
\centering
\caption{Approximate computational context. Deep-model timings are literature-reported GPU timings under different hardware and input settings; ours is measured on CPU.}
\label{tab:timing}
\begin{tabular}{@{}lcl@{}}
\toprule
Method & Reported Speed & Hardware / Source \\
\midrule
ManTra-Net~\cite{mantranet2019} & $\sim$0.8s/img & GTX 1080Ti (1024$\times$768) \\
MVSS-Net~\cite{mvss2022} & 20.1 FPS & Tesla V100 \\
\textbf{Ours} & $\sim$0.4s/img & \textbf{Apple M3 Pro CPU} \\
\bottomrule
\end{tabular}
\end{table}

These timings are not directly comparable due to different hardware, image sizes, and implementations.
The key deployment distinction is that our method runs without GPU acceleration.
On a single CPU core (Apple M3 Pro, 10-core, 16\,GB RAM), our method processes each image in $\sim$0.4s.
With parallelization across 10 cores, batch throughput reaches $\sim$25 images/second.

\subsection{Cross-Validation Stability}

Per-fold results for the format-controlled JPEG-only evaluation demonstrate exceptional stability:

\begin{table}[h]
\centering
\caption{Per-fold AUC for JPEG-only evaluation.}
\label{tab:folds}
\begin{tabular}{@{}cc@{}}
\toprule
Fold & AUC \\
\midrule
1 & 0.991 \\
2 & 0.991 \\
3 & 0.990 \\
4 & 0.990 \\
5 & 0.989 \\
\midrule
\textbf{Mean $\pm$ std} & \textbf{0.990 $\pm$ 0.001} \\
\bottomrule
\end{tabular}
\end{table}

\subsection{Robustness to Post-Processing}

Real-world images may undergo additional processing after manipulation.
We evaluate robustness by applying common degradations before feature extraction on the JPEG-only subset.

\begin{table}[t]
\centering
\caption{Robustness to post-processing degradations (JPEG-only subset).}
\label{tab:robustness}
\begin{tabular}{@{}lccc@{}}
\toprule
Perturbation & AUC & $\Delta$AUC & F1 \\
\midrule
None (baseline) & 0.990 & --- & 0.905 \\
Gaussian blur $\sigma$=1 & 0.962 & $-$0.028 & 0.808 \\
Resize 0.75$\times$ & 0.937 & $-$0.053 & 0.761 \\
Resize 0.50$\times$ & 0.896 & $-$0.094 & 0.689 \\
Gaussian blur $\sigma$=2 & 0.894 & $-$0.096 & 0.672 \\
\bottomrule
\end{tabular}
\end{table}

Performance degrades gracefully under mild perturbations (blur $\sigma$=1: AUC 0.962) and remains above chance even under aggressive 50\% downscaling (AUC 0.896).
This degradation pattern is consistent with a method that genuinely relies on compression artifacts: resampling and heavy blurring destroy fine-grained JPEG quantization traces, which is expected behavior consistent with the proposed detection mechanism.
Note that JPEG recompression robustness is implicitly evaluated via the all-to-JPEG normalization experiments in Table~\ref{tab:main} (q95: 0.952, q85: 0.835, q75: 0.842 AUC).

\subsection{Cross-Dataset Generalization}

To evaluate transfer beyond CASIA v2.0, we train on the full CASIA JPEG-only subset and test without retraining on two external datasets:
(1) the Columbia Image Splicing Detection Evaluation Dataset~\cite{columbia2004} (1,845 fixed-size $128\times128$ image blocks: 933 authentic and 912 spliced), and
(2) a selected JPEG-compressed subset of CoMoFoD small~\cite{comofod2013} (200 base image sets evaluated across 9 JPEG quality variants, yielding 3,600 original/forged test samples).
The CoMoFoD subset is used only for zero-shot transfer evaluation and not for training or threshold selection.

We additionally report a manipulation-type breakdown on CASIA itself (Table~\ref{tab:manip_type}).

\begin{table}[t]
\centering
\caption{CASIA JPEG-only: detection by manipulation type (out-of-fold).}
\label{tab:manip_type}
\begin{tabular}{@{}lcccc@{}}
\toprule
Subset & N\textsubscript{auth} & N\textsubscript{tamp} & AUC & F1 \\
\midrule
All tampered JPEG & 7,437 & 2,064 & 0.990 & 0.905 \\
Splicing only (\texttt{Tp\_S\_}) & 7,437 & 968 & 0.987 & 0.832 \\
Copy-move only (\texttt{Tp\_D\_}) & 7,437 & 1,096 & 0.992 & 0.870 \\
\bottomrule
\end{tabular}
\end{table}

Notably, CASIA copy-move (AUC = 0.992) is detected as strongly as splicing (0.987), indicating that CASIA's copy-move construction process introduces compression-history artifacts detectable by the proposed features.
This contrasts with CoMoFoD, where copy-move preserves compression uniformity (see below).

\begin{table}[t]
\centering
\caption{External transfer and negative-control evaluation: trained on CASIA v2.0 JPEG-only, tested without retraining.}
\label{tab:columbia}
\begin{tabular}{@{}p{2.8cm}llcc@{}}
\toprule
Test Dataset & Type & Role & AUC & F1 \\
\midrule
Columbia~\cite{columbia2004} & Splicing & Transfer & 0.708 & 0.735 \\
CoMoFoD JPEG~\cite{comofod2013} & Copy-move & Neg.\ ctrl & 0.499 & N/A \\
\bottomrule
\end{tabular}
\vspace{2pt}

\footnotesize{CoMoFoD F1 not reported: AUC at chance level; thresholded classification not meaningful under the CASIA-selected threshold.}
\end{table}

\textbf{Columbia (splicing transfer, AUC = 0.708):}
The method achieves above-chance transfer on Columbia's 1,845 $128\times128$ image blocks without retraining.
The moderate result is consistent with the method's design: Columbia blocks lack full-image spatial context and differ from CASIA in scale and compression history.

\textbf{CoMoFoD (copy-move negative control, AUC = 0.499):}
On the JPEG-compressed subset of CoMoFoD small, the method returns chance-level performance.
We interpret this as a negative-control result that delimits the method's scope, not as evidence of copy-move generalization.
CoMoFoD copy-move manipulations copy a region \emph{within the same image}, preserving compression uniformity between source and target regions.
Unlike the cross-source splicing scenario targeted by the proposed features, such manipulations do not introduce heterogeneous double-JPEG responses.
The contrast with CASIA copy-move (AUC = 0.992, Table~\ref{tab:manip_type}) indicates that CASIA's copy-move construction involved editing operations that introduced detectable compression artifacts, whereas CoMoFoD's controlled copy-move protocol preserves compression uniformity by design.

This pair of results delimits the method's operating scope: it detects compression-history inconsistency---effective for cross-source splicing and heterogeneously processed manipulations---but lies outside the strongest operating regime for same-image copy-move where compression history is preserved.
Complementary keypoint- or dense-field-based copy-move detectors are required for the latter scenario.

\section{Discussion}
\label{sec:discussion}

\subsection{Why Multi-Scale ELA Works}

The dramatic improvement from single-quality to multi-scale ELA (+0.180 AUC on the JPEG-only subset) warrants analysis.
JPEG compression is a lossy operation that converges: repeated re-compression at quality $q$ produces diminishing residuals as the image approaches its $q$-specific fixed point.
For an image originally saved at quality $q_{\text{orig}}$:

\begin{itemize}
    \item When $q < q_{\text{orig}}$: Large residuals everywhere due to aggressive re-quantization. Forged regions from a different $q_{\text{forge}}$ converge at different rates.
    \item When $q > q_{\text{orig}}$: Small residuals for authentic content (already below quality ceiling). Forged regions from higher-quality sources exhibit larger residuals.
    \item When $q \approx q_{\text{orig}}$: Near-zero residuals for authentic regions. Double-compressed regions retain quantization table mismatch artifacts.
\end{itemize}

By spanning seven quality levels across $[30, 95]$, at least one level falls in the diagnostic regime for any $q_{\text{orig}}$, resolving the fundamental limitation of single-quality ELA.

\subsection{Interpretability}

Unlike deep learning, our features provide direct forensic interpretation:
\begin{itemize}
    \item High ELA grid variance $\rightarrow$ spatially inconsistent compression history.
    \item Cross-quality ratio anomaly $\rightarrow$ double-JPEG compression signature.
    \item High entropy variance $\rightarrow$ manipulation boundary present.
    \item Abnormal FFT bands $\rightarrow$ resampling or filtering artifacts.
\end{itemize}

\subsection{Limitations}

\textbf{Format dependence.} Our ELA features assume JPEG compression history. Purely lossless workflows (PNG-only) produce uninformative ELA residuals. However, many real-world images are JPEG-compressed at some point in their lifecycle, and the JPEG-only evaluation (AUC 0.990) supports strong performance within this scope.

\textbf{AI-generated content.} Our method targets traditional manipulation (splicing, copy-move) involving recompression. Fully synthetic AI-generated images that have never undergone heterogeneous compression would not produce the double-compression artifacts our features detect.

\textbf{Image-level only.} We provide binary classification without pixel-level localization. The spatial grid features encode positional information but cannot pinpoint exact manipulation boundaries.

\textbf{Robustness ceiling.} Heavy downscaling (0.5$\times$) or strong blur ($\sigma$=2) reduces AUC to $\sim$0.89, which is expected since these operations destroy compression-domain traces. Deployment in scenarios with aggressive post-processing may require complementary features.

\textbf{Cross-dataset scope.} Primary results are demonstrated on CASIA v2.0, with external validation on Columbia (splicing blocks) and CoMoFoD (copy-move negative control). The chance-level CoMoFoD result reflects an intended scope limitation: same-image copy-move can preserve compression uniformity and therefore may not trigger the proposed compression-response features. Broader evaluation on JPEG-sourced forensic benchmarks such as NIST MFC and DSO-1, and complementary copy-move benchmarks such as COVERAGE, remains future work.

\subsection{Reproducibility}

All experiments use: Python 3.13, Pillow 10.4.0 (libjpeg-turbo backend) for JPEG compression/decompression, scikit-learn 1.5.2 for classification, NumPy 1.26.4, SciPy 1.14.1.
JPEG quality parameter maps directly to libjpeg's quantization table scaling.
The HistGradientBoosting classifier uses 255 bins, 1000 max iterations with early stopping (patience 50, 10\% validation fraction), learning rate 0.03, max depth 8, minimum 20 samples per leaf, and L2 regularization 1.0.
All random seeds are fixed at 42.
Feature extraction is deterministic given identical library versions.
Code, evaluation scripts, source-group parsing rules, fold assignments, and out-of-fold prediction files are available at: \url{https://github.com/excitedlord/msela-forgery-screening}.
The repository includes \texttt{casia\_source\_groups.csv}, \texttt{casia\_jpeg\_stratified\_5fold.csv}, \texttt{casia\_jpeg\_grouped\_5fold.csv}, and out-of-fold prediction files for the main controlled protocols.

\subsection{Ethical Considerations}

Image forensics tools can be misused to falsely accuse individuals of tampering or to circumvent legitimate image processing.
Our method should be used as a screening tool within established forensic workflows, not as a sole determinant of image authenticity.
We release evaluation code and feature descriptions to enable independent verification and to support responsible forensic practice.

\section{Conclusion}
\label{sec:conclusion}

We presented a format-controlled evaluation methodology and a GPU-free feature engineering pipeline for image forgery screening.
By identifying and explicitly controlling for the TIFF/JPEG format confound in CASIA v2.0---where a trivial \texttt{is\_tiff} classifier reaches 0.80 AUC---we demonstrate that our multi-scale ELA approach achieves AUC = 0.990 on the format-controlled JPEG-only subset and 0.976 under source-aware splitting based on recoverable CASIA filename identifiers.
The ablation study shows that multi-scale ELA (+0.180 AUC over single-quality on the clean subset) is the dominant mechanism, while the cross-quality ratio features provide complementary double-compression detection from just 16 features.
A manipulation-type breakdown reveals strong detection of both CASIA splicing (0.987) and copy-move (0.992), while the CoMoFoD negative-control experiment (AUC = 0.499 on same-image copy-move) delimits the operating scope: the method detects compression-history inconsistency introduced by heterogeneous processing, not intra-image duplication that preserves compression uniformity.

Our results establish two conclusions: (1) carefully engineered JPEG compression response features provide strong image-level compression-history screening with feature-level interpretability and CPU-only deployment; (2) format-controlled evaluation is essential when working with CASIA v2.0---prior results reported without such controls may be susceptible to format-related inflation.
Overall, the proposed method should be used as a CPU-only screening tool for JPEG compression-history mismatch, especially in cross-source or heterogeneously processed manipulations, rather than as a standalone universal forgery detector.

Future work includes: (1) pixel-level localization using grid feature gradients, (2) cross-dataset evaluation on NIST MFC and modern JPEG-sourced splicing benchmarks to better characterize transfer under matched compression conditions, (3) extension to AI-generated content detection, and (4) complementary copy-move detection via keypoint matching or dense field estimation.


\end{document}